# Insights to Molecular and Bulk Mechanical Properties of Glassy Carbon Through Molecular Dynamics Simulation and Mechanical Tensile Testing

Manali Kunte[a,b+], Lucía Carballo Chanfón[a,b+], Surabhi Nimbalkar[a,b], James Bunnell[a,b], Emanuel Rodriguez Barajas[a,b], Mario Enrique Vazquez[a,b], David Trejo-Rodriguez[a,b], Carter Faucher[a,b], Skelly Smith[a,b], Sam Kassegne[1,a,b]

[a]NanoFAB.SDSU Research Lab, Department of Mechanical Engineering, College of Engineering, San Diego State University, San Diego, CA, USA 92182, [b]NSF-ERC Center for Neurotechnology

[+]*Manali Kunte and Lucía Carballo Chanfón contributed equally to this work.*

**ABSTRACT**
With increasing interest in the use of glassy carbon (GC) for a wide variety of application areas, the need for developing fundamental understanding of its mechanical properties has come to the forefront. Further, recent theoretical and modeling works that shed some light on the synthesis of GC through the process of pyrolysis of polymer precursors have highlighted the possibilities of a revisit to investigation of its mechanical properties at a fundamental level. While there are isolated reports on the experimental determination of its elastic modulus, insights into stress-strain behavior of GC material under tension and compression obtained through simulation, either at molecular level or for the bulk material is missing. This current study fills the gap at the molecular level and investigates the mechanical properties of GC using molecular dynamics (MD) simulations which model the atomistic level formation and breaking of bonds using bond-order based reactive force field formulations. The molecular model considered for this simulation has a characteristics 3D cagey structure of *5-, 6-, and 7-membered* carbon rings and graphitic domain of a flat graphene-like structure. The GC molecular model was subjected to loading under varying strain rates (*0.4/ns, 0.6/ns, 1.25/ns, and 2.5/ns*) and varying temperatures (*300 - 800 K*) in each of the three axes *x, y, and z*. The simulation showed that GC molecule has distinct stress-strain curves under tension and compression. In tension, MD modeling predicted mean elastic modulus of *5.71 GPa* for a single GC molecule with some dependency on strain rates and temperature, while in compression, the elastic modulus was also found to depend on the strain rates as well as temperature and was predicted to have a mean value of *35 GPa*. For validation of the simulation results and developing an experimental insight into the bulk behavior, mechanical tests were carried out on *dog-bone* shaped testing coupons that were subjected to uniaxial tension and loaded until failure. The GC test coupons demonstrated a *bulk modulus* of *17 ± 2.69 GPa* in tension which compare well with those reported in the literature. However, comparing MD simulation outcomes to those of uniaxial mechanical testing, it was clear that the *bulk modulus* of GC in tension found experimentally is higher than modulus of single GC molecule predicted by MD modeling, which inherently underestimate the *bulk modulus*. With regard to failure modes, while the MD simulations predicted failure in tension accompanied by breaking of carbon rings within the molecular structure, the mechanical testing, on the other hand, demonstrated failure modes dominated by brittle failure planes largely because of the amorphous structure of GC.

---

[1] Address correspondences to Sam Kassegne • Professor of Mechanical Engineering, NanoFAB.SDSU Research Lab, Department of Mechanical Engineering, College of Engineering, San Diego State University, 5500 Campanile Drive, CA 92182-1323. E-mail: kassegne@sdsu.edu • Tel: (760) 402-7162.

## 1. INTRODUCTION

The past several decades have seen increasing level of interest in the use of glassy carbon (GC) material, formed through pyrolysis of photo-patternable epoxy resins, in a wide variety of application areas such as energy storage, biochemical sensing, neural probes, nanoelectronics, and nanofabrication, among others [1-8]. This has been driven mainly by the superior and well-documented electrical, electrochemical, and chemo-thermal properties that GC offers such as chemical stability, thermal resistance, good electrical conductivity, wide electrochemical window, etc. [7-8]. While compelling mechanical properties have not particularly been a critical requirement for the majority of the application areas explored so far, properties such as elastic modulus, yield and ultimate stress, and failure modes will become even more important and relevant as GC's application areas grow.

Related to this, developing a comprehensive understanding of the precise molecular nature of GC has, of course, a large bearing on further gaining insights to its bulk and molecular mechanical properties. However, despite the widespread use of GC, understanding of the process of formation of its molecular structure during pyrolysis stages, and hence its mechanical properties at the molecule level continue to evolve, especially with emerging new insights from molecular dynamics modeling [9-10]. The molecular models of GC proposed by Franklin [11], Harris [12-14], and Harris and Tsang [15] consider GC to consist of interconnected graphene ribbons, whereas the model of Jenkins and Kawamura emphasize a cage-like graphene structure in addition to crosslinked ribbon-like structures created due to the bonds formed between different functional groups [16].

Recent modeling efforts have also provided fresh look at the molecular structure of GC. Notable examples include recent work of Tomas *et al* in modeling the graphitization of amorphous carbon and topology of disordered graphene networks and a recent report on structural transformation of GC under compression [17-20]. Current understanding of the molecular structure of GC supported by MD modeling, therefore, suggests that, as a bulk material, GC has $sp^2$ and $sp^3$ hybridization states and predominantly consists of a flat graphene-like $sp^2$ hybridized structures together with 3D cage-like component of *5-, 6-, and 7-membered* carbon rings that are in turn attached to neighboring carbon structures through $sp^3$ hybridization [9-16, 21].

With regard the mechanical behavior specifically, there are few experimental works reported in the literature. Jurkiewicz *et al* reported that elastic modulus of GC ranges from *15.6 GPa - 37.6 GPa* [21] whereas Kotlensky and Fischback reported an elastic modulus of *~24 GPa* in tension [22]. Our group had investigated variation of elastic modulus of GC microelectrodes with varying ramping rates (2.32 °C/min - 8.1 °C/min) and holding temperatures (600 – 1000° C) of the pyrolysis process and reported elastic modulus of *20 GPa – 55 GPa*, with the maximum at 800°C and rapidly decreasing thereafter [5]. From the theoretical perspective, molecular dynamics simulation has been used to explore the molecular-level mechanical properties of various carboneous materials such as diamond, graphene, and CNT [23], amorphous carbon [24], multi-layer graphene oxide [25], and epoxy resin molecules [26].

However, to-date, there is no reported work on the use of MD simulation in predicting the molecular-level mechanical properties of GC, particularly its stress-strain behavior and elastic modulus. To address this gap, we explore here the molecular-level mechanical properties of glassy carbon material through reactive molecular dynamics formulations using the *ReaxFF* Software. Variation in the direction of strain, ambient temperature of the molecule, and strain rate applied were taken into consideration. For model validation and comparison of the stress-strain behavior and the elastic modulus of the bulk GC material, we also microfabricated GC test structures and carried out uniaxial tensile load testing.

## 2. MATERIALS AND METHODS

### 2.1 Molecular Dynamics Simulations
In this work, reactive molecular dynamics (MD) simulations using bond-order based reactive force field was performed using the ReaxFF software within *SCM's Amsterdam Modeling Suite (AMS)* [27-30]. Briefly, ReaxFF is a relatively fast MD simulation tool that is based on empirical interatomic potentials together with force-fields derived from quantum mechanics structure and energy data and is specifically developed to capture chemical changes due to reactions and external forces. The force-fields are functions of atomic positions and include terms associated with non-reactive interactions such as electrostatic, van der Waals, and angle-strains as well as reactive terms arising from connection-dependent terms that model changes in atom connectivity [27-30]. MD simulation allows for accurate description of bond dissociation, bond formation, and exploration of reaction pathways since the formulation is based on *bond-order* and *bond-distance* and *bond-order* and *bond-energy* relationships. Overall, since first-principle based approaches are computationally expensive, the empirical *bond-order* potentials of ReaxFF offer an acceptably accurate tool.

As shown in Figure 1, the starting model of undeformed GC molecular structure is the molecule already obtained in our prior work and consists of 1322 atoms (mainly carbon with some oxygen and hydrogen present) arranged in a cage-like component in the form of *5-, 6- or 7-membered rings* and a flat graphitic domain consisting of graphene-like carbon rings in a simulation cell of *34 Å x 34.1 Å x 33.9 Å* [9]. There are approximately 70, 130, and 140 of *5-, 6- and 7-member* rings respectively in this molecule [9].

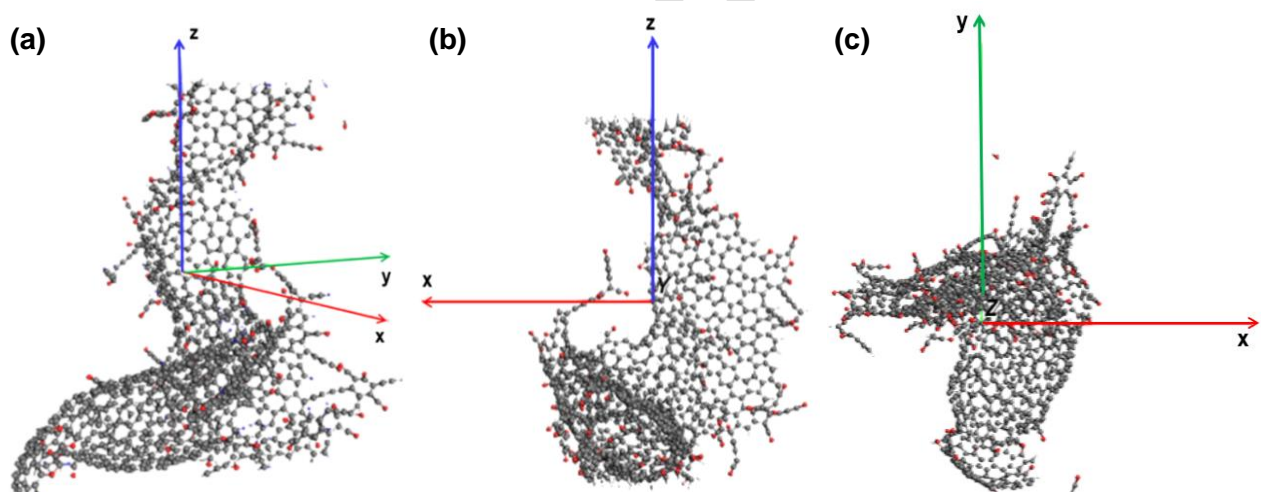

**Figure 1.** (a) molecular structure of glassy carbon (b) view along the *y-axis* showing flat planar structure along the *x-z* plane (c) view along *z-axis* showing cage-like nanostructure along *x-y* plane.

### 2.2 Assumptions, Boundary Conditions, and Parameters for MD Model
All models were analyzed at constant temperature (300K) using a temperature damping constant of 100 fs and a pressure damping constant of 500 fs with constant volume and temperature conditions through the constant temperature and volume (NVT) Berendsen thermostat [30]. Cell parameters were kept constant while using the thermostat. Time step of 0.25 fs for total run time of 2000 fs was used. To determine the effect of direction of loading *vis-a-vis* anisotropic material property of GC, three separate loading cases were considered in each of the x, y, and z axes under a periodic boundary condition. The choice of applying loads (strain rates) in the 3-separate axis informs the effect of anisotropy on the material properties. Further, separate simulations were carried out for uniaxial tensile and compressive strains applied in each of the *x, y, and z* direction for a variety of strain rates (*0.4/ns, 0.6/ns, 1.25/ns, and 2.5/ns* for tension and *1.5/ns, 2.5/ns* for compression) and temperature ranges (*300 - 800 K*). These selected strain levels were obtained through optimization runs that identified strain rates that produced adequate sampling of data points. Figure 2 illustrates one of the directions in which the strain loads were applied. Once the simulations were run, data in terms of the energy, absolute strain and stress were extracted from the raw results file using the open-source python code '*stress_strain_curve.py*' [26]. The data is then compiled and plotted to obtain the stress-strain curve, from which the elastic modulus was obtained through curve-fitting its linear region.

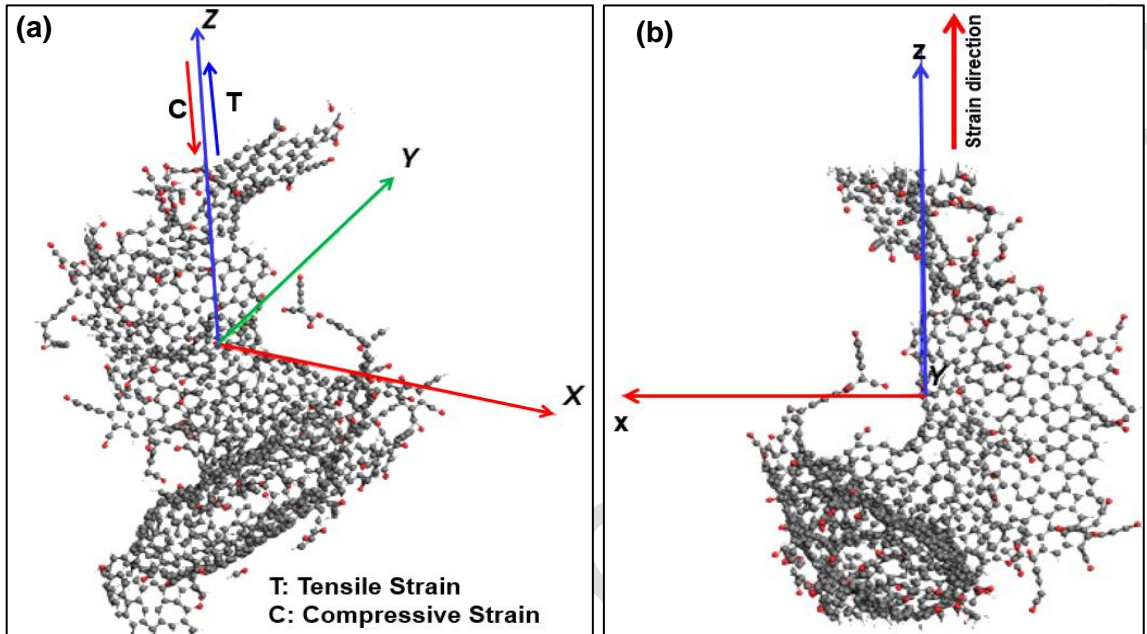

**Figure 2. (a)** directions of strain applied to GC molecule (isometric view) **(b)** tensile strain shown applied along the *z-axis*. Strain rates of *0.4/ns, 0.6/ns, 1.25/ns, and 2.5/ns* were used.

**2.3 Mechanical Test Through Uniaxial Tensile Loading**
*Dog-bone* shaped test structures (coupons) were used for uniaxial tensile loading experiments. Dimensions of the *dog-bone* test coupons were determined following the recommendation of *ISO II (DIN 20*80) standards* for tensile testing of metals and subsequently scaled down by a factor of 5 due to the constraints of *4" wafer* microfabrication [31]. The final geometry adopted is shown in Figure 3 with dimensions of widths of *6 mm* at its base and *3 mm* in the neck region and total length of *33 mm* that further consists of *8.5 mm* long component in the main section and *8 mm* in the neck region. The eight GC test structures were microfabricated following well-known processes of negative tone lithography and pyrolysis, as described elsewhere [6-7]. Out of these eight specimens, four consisted of plain GC structures that were lifted off the substrate after pyrolysis using buffered hydrofluoric (BHF) acid, while the other four were supported on a flexible polyimide substrate and then released [32-33]. Instron 1500HDX Universal Machine (Instron, USA) was then used to characterize these specimens under uniaxial tensile loading. Here, linear displacement due to the crosshead motion of Instron was used to obtain displacement values. Thread-rolled C-clamps with clamping capacity of *0-25 mm* were held between the jaws of upper and lower clamps of the Instron machine respectively. The crosshead extension rate was set at *0.1 mm/min* and the specimens were loaded to failure. After the testing coupon structure failed, the resulting extension was measured. The stress-strain graphs were plotted and the elastic modulus of the eight samples was calculated by taking the average of ten points on the graph of each sample. FASTCAM SA-Z high speed camera (Photron, USA) was used for capturing images during loading and at failures.

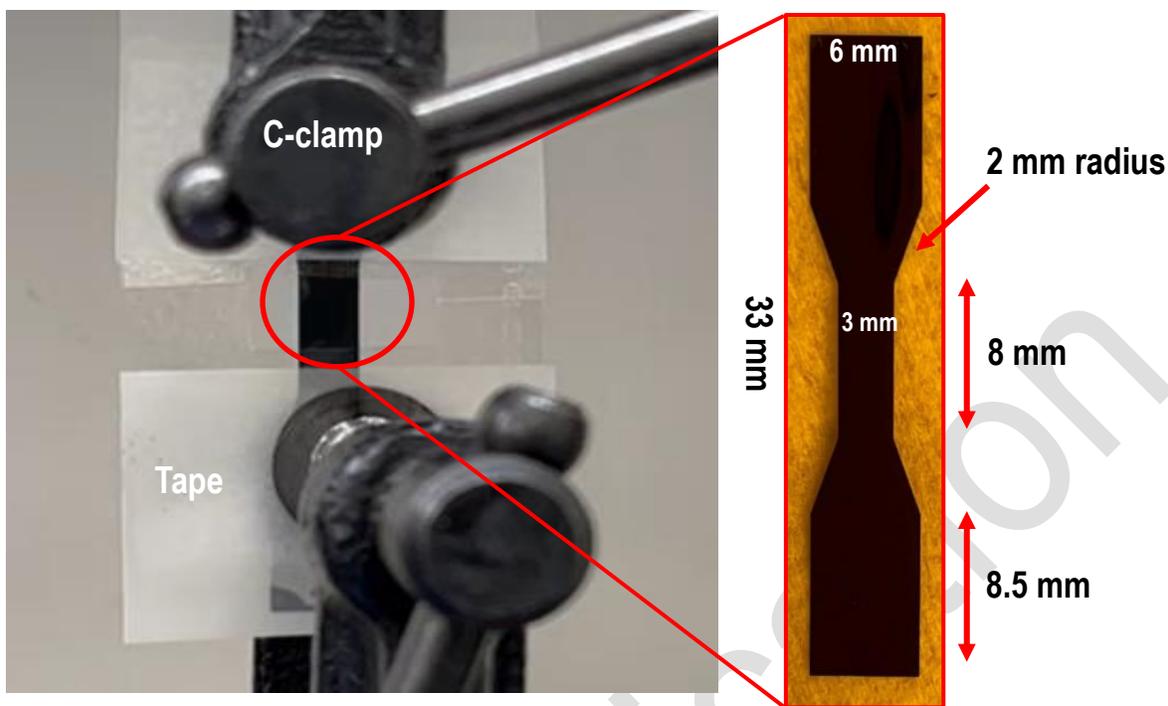

**Figure 3.** Experimental set-up for uniaxial tensile testing of GC *dog-bone* shaped test structure (dimensions shown) using the Instron 1500HDX Universal Machine. FASTCAM SA-Z high speed camera was used for image and video capture. Geometry of the coupons used is indicated.

## 3. RESULTS
The results highlighting predictions of stress-strain curves as well as elastic modulus of the GC molecule under a variety of strain rates and temperature are presented.

### 3.1 Stress-Strain Relationships and Modes of Failure of GC Molecule Under Tension
Figure 4 shows the progression of failure of the GC molecule at different time points in the MD simulation (strain rate = *1.25/ns*). As the strain level increases, localized dislocation at sites in the flat graphitic component of the molecule that has fewer number of carbon rings is observed. This localized stretching of ring structures continues until strain level of *0.175 Å/Å* is reached at which point the first carbon ring structure breaks as shown in Figure 4e (e.g., for strain rate of *1.25/ns*). Figure 5 shows a close-up of this first failure point in a *10–member* ring structure in which the bond between carbon molecules designated as $C_{747}$ and $C_{1128}$ atoms breaks. The GC molecule continues to support the applied strain with new interlocking of loose chains generated until the next set of bonds between carbon atoms designated as $C_{600}$ and $C_{819}$ breaks at strain level of *0.195 Å/Å* (for strain rate of *1.25/ns*). Transient plot of energy of the system and distance between carbon atoms as shown in Figure 6 offers another insight to these breakings of bonds between carbon atoms in a ring structure. The locations of these failures in carbon bonds in the rings correspond to energy peaks. For example, the first energy peak that corresponds with the first failure is accompanied by a corresponding increase in the interatomic distance between $C_{747}$ and $C_{1128}$ atoms from ~2Å to ~15Å. Similarly, at the second energy peak, the interatomic distance between atoms $C_{600}$ and $C_{819}$ atoms increases from ~2Å to ~5.9Å, corroborating the observation at the second failure point.

Regarding the stress-strain relationships, Figure 7 shows that until the first failure at strain levels of *0.175 – 0.18 Å/Å*, linear elastic behavior is dominant for all strain rates considered. Further, loading through slower strain rates appears to lead to modest decreases to the strain corresponding to the ultimate stress ($\sigma_{ult}$) even though the difference in the actual ultimate stress is negligible. For example, the ultimate stress corresponding to strain rate of *0.4/ns* is 1 MPa, while for *2.5/ns,* it is 0.95 MPa – barely 5% lower. For all strain rates -- subsequent to the first failure and the accompanying reduced load carrying capacity -- secondary and sometimes tertiary (for example for strain rate of *1.25/ns*) re-loading paths were observed that eventually led to complete failure.

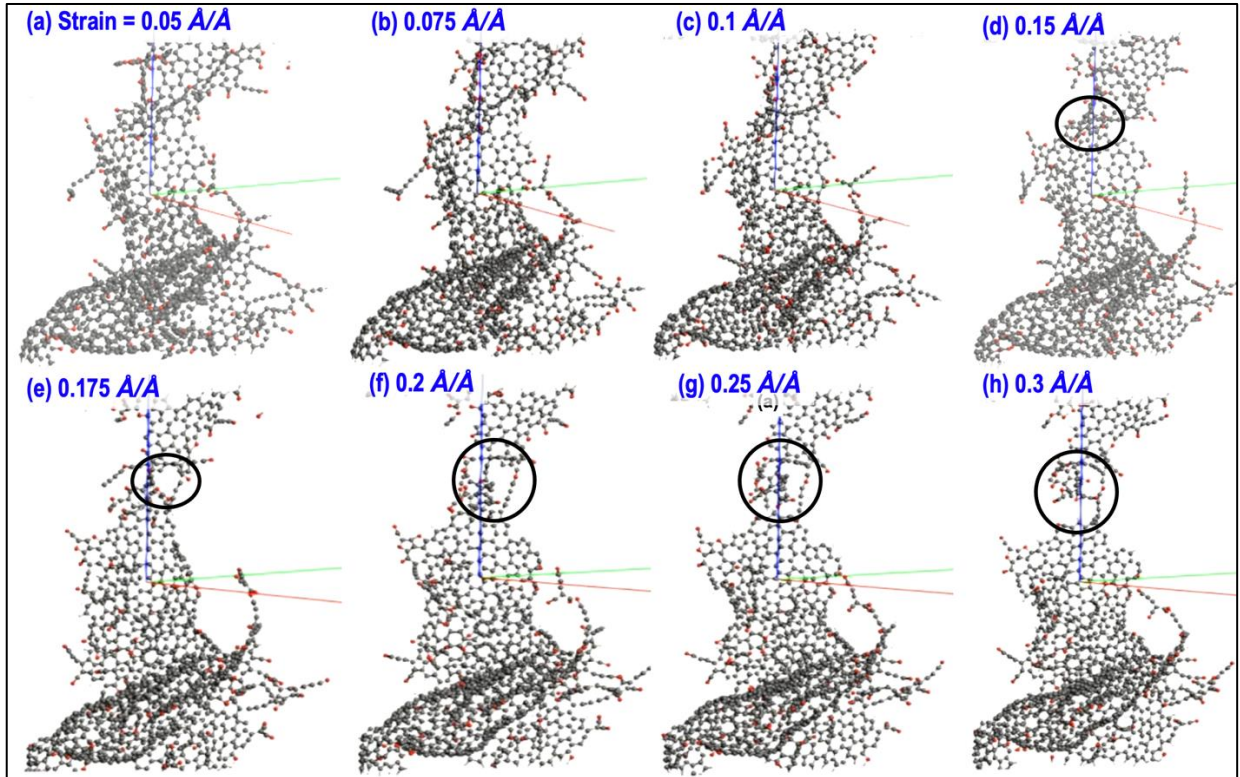

**Figure 4.** Progression of failure of GC molecule with increase in time and loading through strain (strain rate = *1.25/ns*). Breakage of $sp^2$ bonds eventually led to *c-c* chains before complete failure.

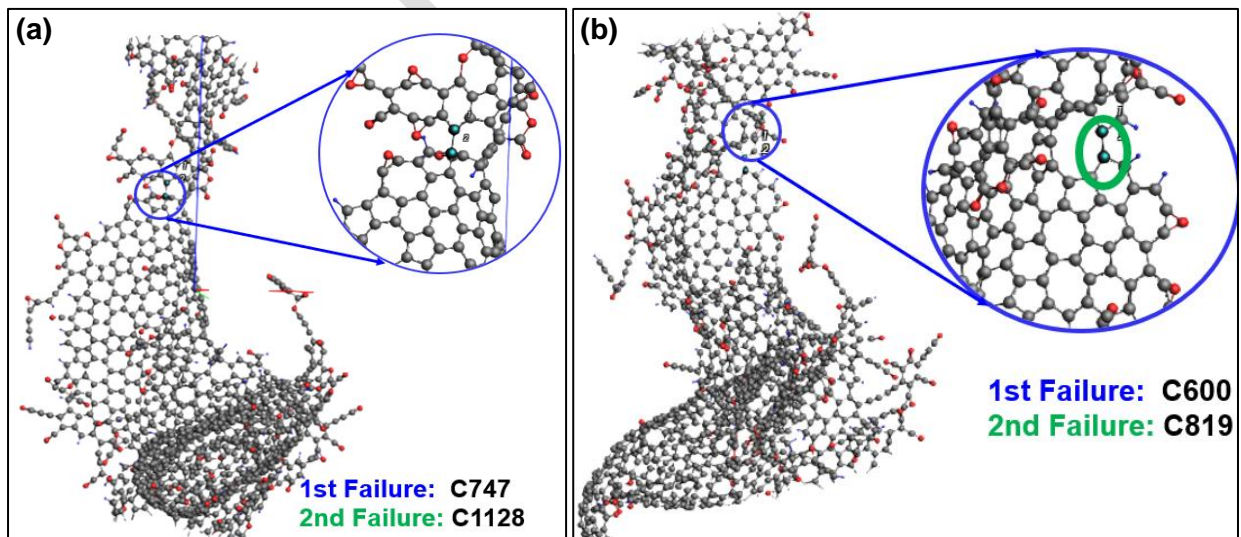

**Figure 5.** (a) bonds in *10–member* ring structure break under tensile load along the *z-axis*. It is followed by various link breakages in the adjoining structure (b) Bond that breaks corresponding to the second energy peak under tensile load along the *z-axis*.

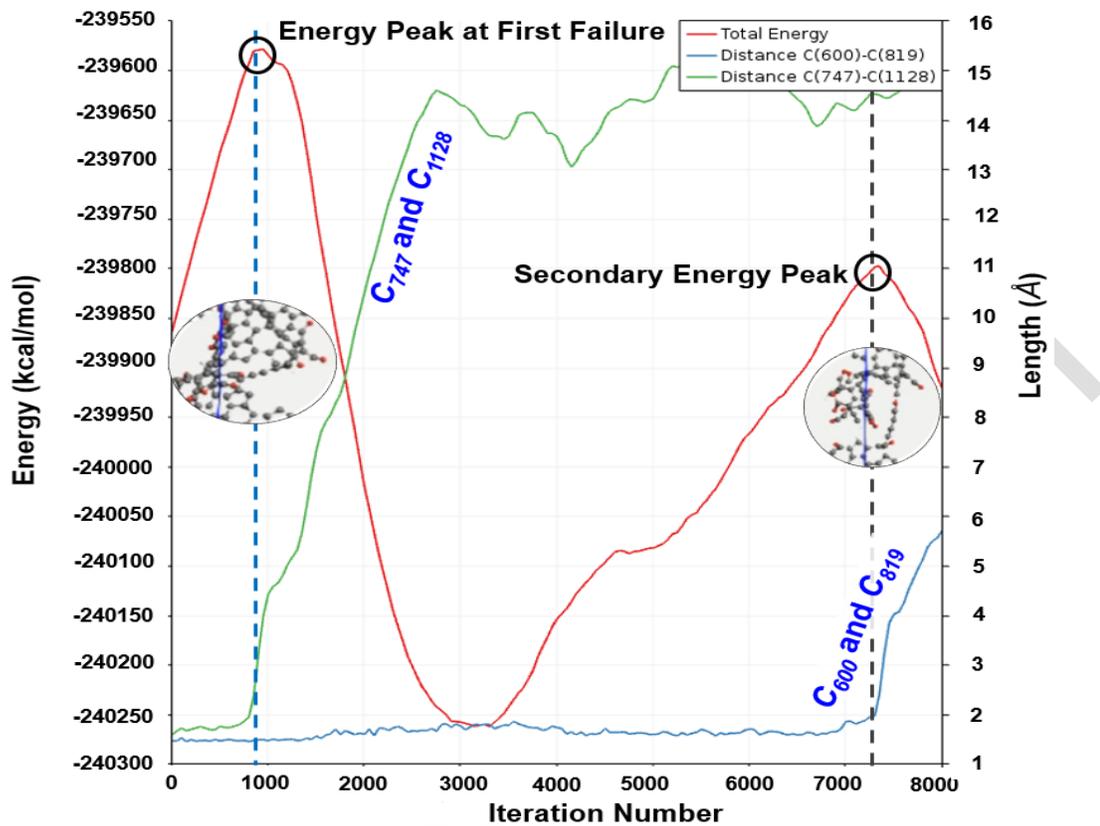

**Figure 6.** Energy plot and plot of distance between the two sets of atoms $C_{747}$ and $C_{1128}$ and $C_{600}$ and $C_{819}$ corresponding to the two energy peaks. 8000 iterations correspond to 2000 fs, total run time.

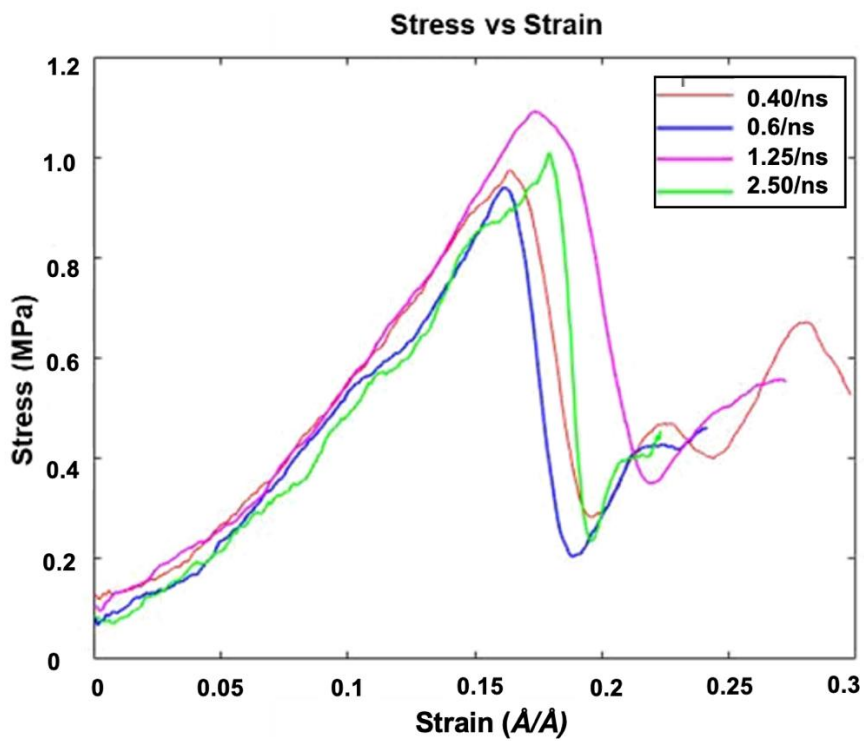

**Figure 7.** Stress-strain diagram for strain rates of *0.4/ns, 0.6/ns, 1.25/ns, and 2.5/ns.* Low strain loadings allow enough time for stresses to respond to deformation resulting in higher failure strains.

## 3.3 Effect of Temperature on Stress-Strain Behavior

To explore the effect of temperature on the elastic modulus of GC molecule, constant strain rate of *0.6/sec* and constant pressure of *1 atm* were applied to the GC molecule as the temperature was varied from *300 K – 800 K*. Figure 8a summarizes the stress-strain curves under these temperature ranges and shows elastic modulus values in the narrow range of *5.71 ± 0.23 GPa.* Further, Figure 8a highlights that – in the elastic region - the stress-strain curves corresponding to the temperatures range considered here, in general, exhibit similar initial slope and hence similar elastic modulus, with the mean elastic modulus of *5.71 GPa*. Slight decrease in the ultimate stress in the range of *<5%* was observed between 300 K and 800 K cases. However, in the post-elastic region past the first peak, the stress-strain curves diverge significantly highlighting the effect of temperature. In this region, a rise in temperature is observed to result in early occurrence of second failure strain levels accompanied by a reduced corresponding stress.

The energy vs time plot of Figure 8b shows that after the first failure, the energy of the system drops slightly *(<1%)* for each temperature considered and follows a secondary path until failure. At the relatively lower temperatures, a peculiar behavior of a flat region of the energy indicates a creep-like behavior where increasing strain level is accompanied by a lagging increase in stress. It is also interesting to note that the relationship between temperature and onset of first failure is nonlinear where, for example, for 800 K case, there is >*10%* delay in time of first failure compared to the 300 K case. However, the failure modes of the GC molecule remain to be the same for all temperature ranges considered indicating that failure is a consequence of the loading exceeding the force of existing bond between carbon atoms with temperature effects being secondary.

## 3.4 Stress-Strain Relationships of GC Molecule under Compression

To analyze the elastic modulus under compression, the model was loaded with compressive strains rates of *1.5/ns* and *2.5/ns* in the negative direction along the *z–axis*. These strain rates that are moderately higher than those in the tensile model were selected to ensure that adequate datapoints were obtained for a reasonable stress-strain curve, as lower strain levels resulted in fewer data points. Figure 9 summarizes the progression of failure under compression loads along *z-axis*, where the flat part of GC molecule experiences compressive deformations until it is practically flattened on the 3D cagey component, at which point the load carrying capacity actually increases significantly.

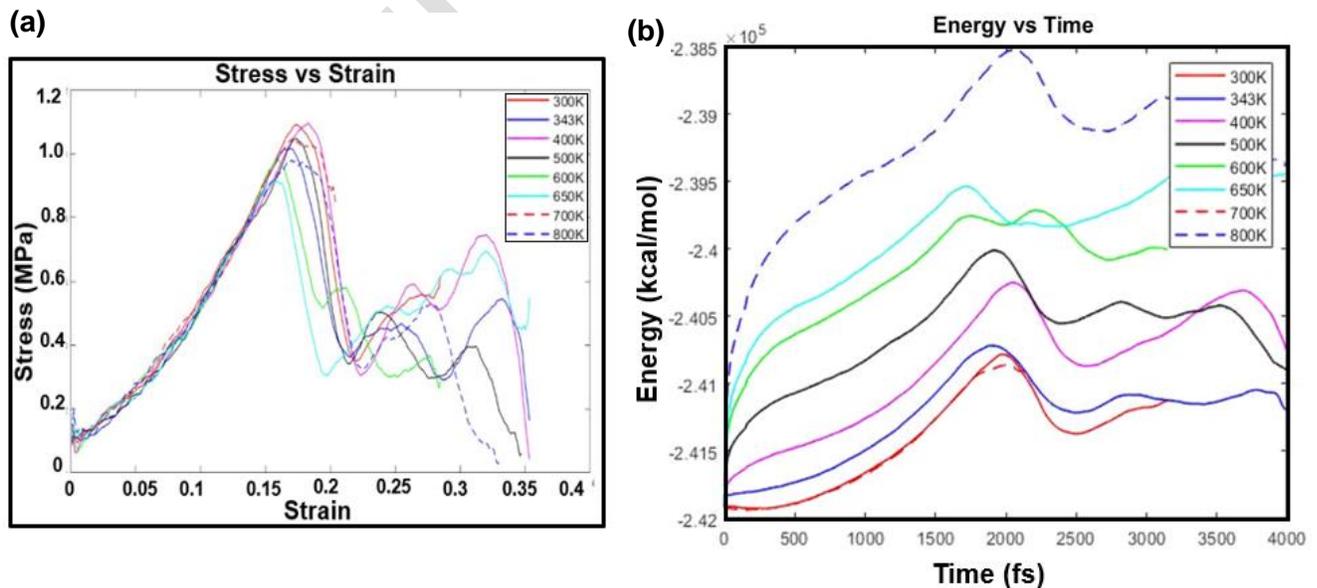

**Figure 8.** (a) stress-strain curves (b) energy vs time curves for *300 K - 800 K* temperature range.

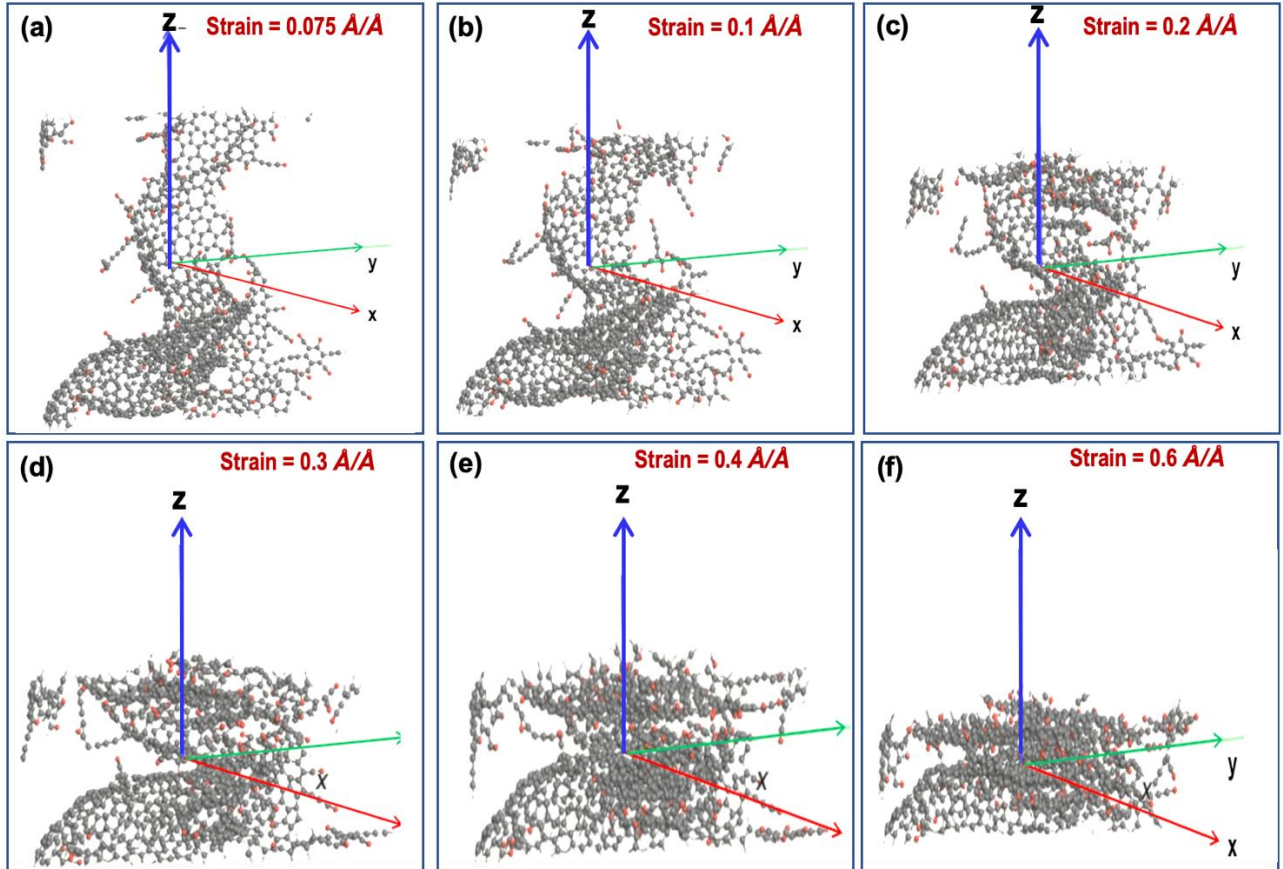

**Figure 9.** *Deformation of* GC molecular structure at different points of time under compression along the *z-axis* (strain rate = *1.5/ns)*.

## 3.5 Experimental Results for Uniaxial Tensile Load Test

A total of eight samples were tested at a loading rate of *0.1 mm/min*, with four plain and bare GC samples and additional four GC samples supported on a flexible polyimide substrate. The GC coupons supported on polyimide substrates were used to evaluate the effect of such substrates on the stress-strain behavior as well as failure modes of GC microstructures. All the bare GC samples were able to withstand a similar level of maximum load of *1.9 – 2.8 N* and a similar extension varying from *0.1 – 0.3 mm*. As shown in Figure 10a, the stress-strain curves of bare GC structures demonstrate a brittle failure with a mean maximum stress ($\sigma_{ult}$) of 81.5 MPa and corresponding mean strain at failure of 0.825%. The elastic modulus was found to be *17.0±2.69 GPa*. Consistent with brittle type of materials, the failure occurred as the ultimate stress with no distinct yielding. For the GC supported on polyimide substrate (Figure 10b), the mean elastic modulus was found to be *3.3 ± 0.59 GPa,* with a mean maximum stress ($\sigma_{ult}$) of 65.5 MPa and a corresponding mean strain at failure of 2.15%. The failure behavior of both specimen of GC is summarized in Figure 11. For the case of bare GC, the failure was typically initiated at the corners of the neck region and progressed fast resulting in large chunks that tore of the main structure along with a typical brittle-type failure consisting of long pieces. An interesting aspect is that the first failure mode was always primarily tearing at corners followed by brittle failures emanating from the torn pieces as well as parts of the neck region of the *dog-bone*. Additionally, all the fractures tend to be located in the same area, the upper and lower neck of the sample. For the GC structures supported on a polyimide surface, the failure mode shown in Figure 11b demonstrates a clean tearing type of failure in the neck region, that is consistent with failure types in ductile materials. The polyimide layer provided this ductility that compensated for the tendency for brittle failure of the GC layer.

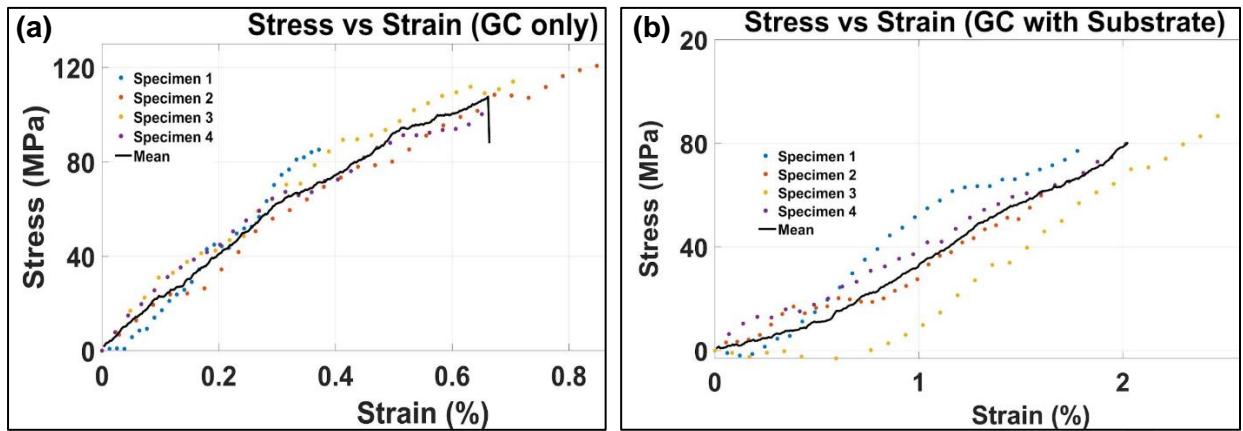

**Figure 10.** stress-strain curve obtained experimentally for (a) plain GC *dog-bone* coupons and (b) GC coupons supported on polyimide substrate. The elastic modulus was obtained through curve-fitting of the linear region.

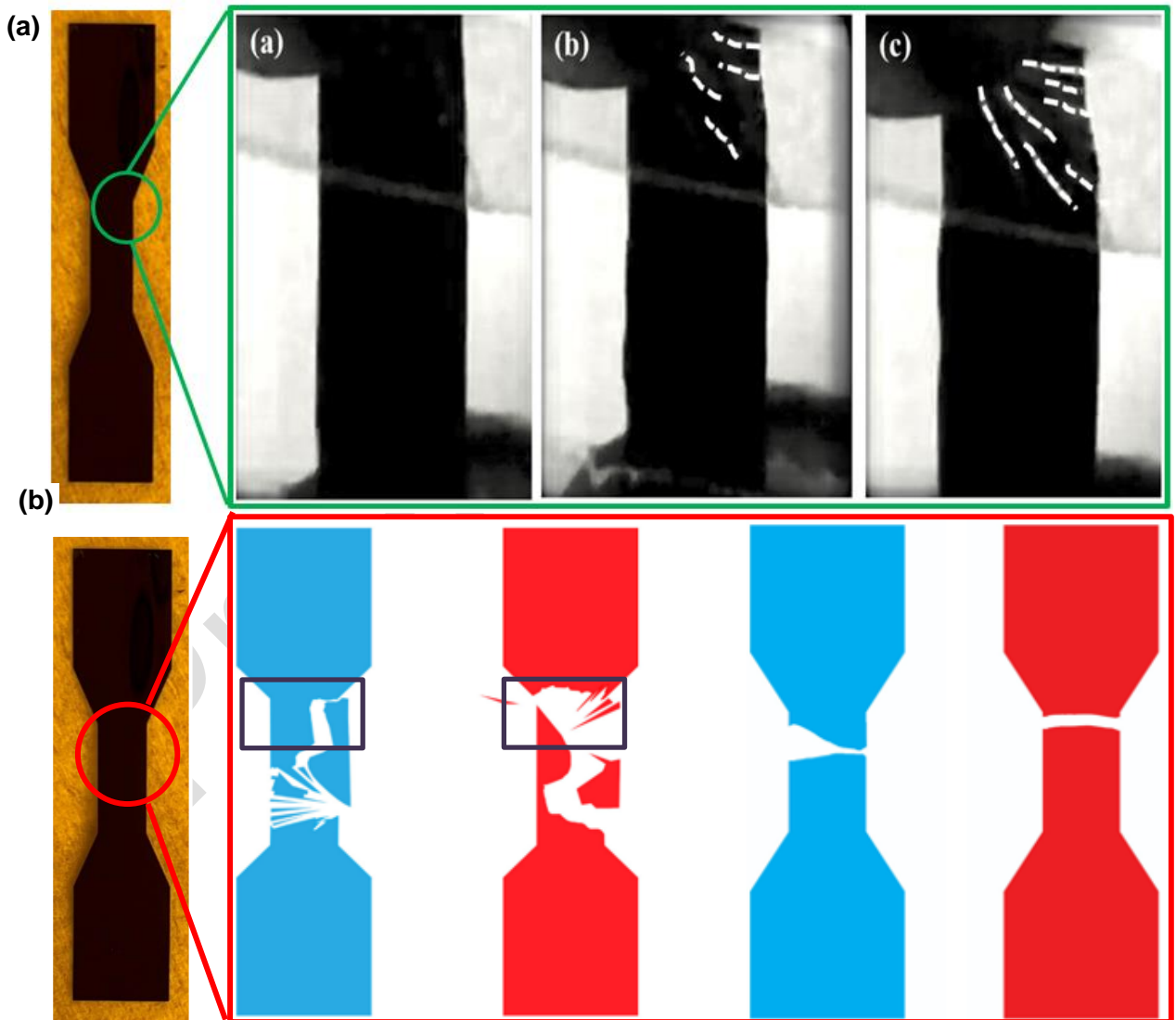

**Figure 11.** (a) failure propagation in test coupon through cracks (b) rendering of failure modes in tension for bare GC test structures and GC test structures with polymer substrate.

## 4. DISCUSSIONS
Both the MD modeling and experimental uniaxial testing provide some unique insights to the mechanical behavior of GC. The MD modeling of GC molecule clearly demonstrate key characteristics such as anisotropy, different axial and compressive load responses, multiple failure modes, and some dependency on temperature and strain rates that are discussed in detail below.

### 4.1 Stress-Strain Response and Failure Modes in Tension vs Compression
The MD modeling has highlighted that the stress-strain responses under tension and compression are quite different from each other. For example, failure in tensile loading of GC molecules happens due to the tearing or breaking of weak links in the flat graphene-like $sp^2$ hybridized components where the interatomic forces due to tensile load exceeds the bond forces, whereas in compression, the molecule undergoes continued compaction that eventually results in a '*snap-in*' or crashing phenomenon that allows it to carry even higher loads. This is, interestingly, consistent with the model predicted by R. Franklin on non-graphitizing allotropes such as GC [11]. In general, because of the absence of grain boundary in GC molecule, failures are sudden and occur along multiple locations wherever there is a progression of thinning of strength due to breaking bonds along weak lines. The sudden, brittle, and fast progressing nature of typical failures in this and similar materials can be explained by the fact that the breakage of a bond in $sp^2$ type components leads to the transfer of load to adjacent atomic bonds, which in turn will lead to stress concentration in the new locations and a subsequent crack growth [23-24, 34].

### 4.2 Effect of Temperature and Strain Rates on Response
In general, the effect of temperature on the stress-strain behavior in the temperature ranges considered (*300 K – 800 K*) was relatively small. For example, as temperature was raised, the maximum stress ($\sigma_{ult}$) experienced decrease in the range of *<5%*. However, a significant behavioral effect was observed in the post-yield region where the stress-strain curves diverge noticeably highlighting the effect of temperature. In this region, with an increase in temperature, secondary failure points occurred with decreasing strain levels accompanied by reduced corresponding stresses. This may be due to the different level of active thermal energy available for rearranging the molecular structures post-failure. Similarly, the effect of loading strain rates was also modest where loading through slower strain rates appears to lead to modest decreases to the strain corresponding to the ultimate stress even though the difference in the actual ultimate stress is negligible. For example, in the case of loading through strain rate of *0.4/ns*, the ultimate stress was 1 MPa, while for *2.5/ns*, it is 0.95 MPa; barely *5%* lower. More importantly, however, for all strain rates, subsequent to the first failure and the accompanying reduced load carrying capacity, secondary and sometimes tertiary (strain rate of *1.25/ns*) re-loading paths were observed that eventually led to complete failure. This is not unexpected as failure was progressive and happened after the weakest links failed first followed by redistribution of loads to the remaining rings that eventually failed as the loading continued.

### 4.3 Anisotropic Behavior
GC molecule has a non-crystalline and amorphous structure that gives rise to anisotropic mechanical behaviors. This anisotropy could partially explain the non-insignificant variations in the value of modulus of GC reported in the literature, as test structures built do not have any fixed direction of orientation of network of GC molecules in the bulk material. The anisotropy, of course, contributes to difficulty in identifying and choosing loading planes to determine stress-strain behavior. In typical cases, a load applied to such a molecule in the axes tagged here as *x*- and *y*- axes produces additional responses such as shear which obviously complicate the separation of single axis stress-strain responses. In addition, from the modeling perspective, limitations in MD modeling in assigning a neutral axis complicated evaluation of moduli in the *x* and *y* directions, as the model produced rotations and rigid body motion. Nonetheless, the anisotropy was obvious through visual inspection of the GC molecular model.

### 4.4 Molecular Behavior Compared to Bulk Behavior
The MD model used here considered only a single molecule and, therefore, does not fully represent the behavior of the bulk GC material. In the bulk material, these molecules interconnect through $sp$, $sp^2$, and $sp^3$ hybridization states and form a complex anisotropic structure with all its impurities and defects along with pores. Therefore, while a model based on a single molecule is useful in offering insights into important phenomenon such as potential failure modes, effect of loading directions, temperature, and strain rates, it cannot, however, predict the influence of neighboring molecules in the bulk response. Further, the bulk behavior also represents the averaged response of all the molecules in the bulk material that consists of complex structures. This basic difference between the property of a single molecule and the bulk material, therefore, accounts for the major part of the difference in the modulus predicted by MD model and that found experimentally through uniaxial testing.

### 4.5 Comparison of Elastic Modulus Values of GC Reported in Literature
Table 1 summarizes some of the modulus values for GC bulk material reported in the literature that are in a range between a low of ~15 GPa to a high of 55 and 62 GPa, demonstrating a rather significant variation among these values. In the majority of the reported cases, the moduli were determined through nanoindentation experiments and, therefore, reflect the mechanical property of GC under compression or flexure [5, 21, 37-39]. Out of the remaining reported cases that were done through a direct tensile testing [22,36,40], the work of Kotlensky *et al* [22] is based on GC material that was heated to almost 2600 °C and then annealed at still high temperatures making its direct comparison with others rather difficult to justify. Therefore, in general, comparison of values of modulus reported here with those from the literature requires the consideration of the not only the difference between bulk and molecular response but also a recognition of the difference of behavior under tension and compression.

| **Source / Reference** | **Bare GC/Substrate** | **Type of Load** | **Modeling/Test** | **Modulus (GPa)** |
|---|---|---|---|---|
| Current | Bare GC | Tension | MD Modeling | 5.71 |
| | Bare GC | Compression | MD Modeling | 35 |
| | Bare GC | Tension | Uniaxial Load | 17.0 ± 2.69 |
| | Polyimide Substrate | Tension | Uniaxial Load | 3.3 ± 0.59 |
| Nimbalkar *et al* [7] | Polyimide Substrate | Tension | Uniaxial Load | 2.65 |
| Kassegne *et al* [5] | Bare GC (700 um φ pillars) | Compression | Nanoindenter | 20 – 55 |
| Jurkiewicz1 *et al* [21] | Bare GC thin-film from furfuryl alcohol as a precursor | Compression | Nanoindenter | 15.6 – 37.6 |
| Kotlensky *et al* [22] | Bare GC (*dog-bone*) | Tension | Uniaxial Load | 14.5 – 28.3 |
| Jenkins & Kawamura [36] | Bare GC | Tension | Uniaxial Load | 30 |
| Yang, Dang, & Ruan [37] | Bare GC | Flexure | Flexural Vibration (IET) | 21.5 |
| Field and Swain [38] | Bare GC | Compression | Spherical Indenters | 20.8 |
| Albiez and Schwaiger [39] | Bare GC | Compression | Nanoindenter | 27-47 |
| Manoharan *et al* [40] | Bare GC – Thin Film | Tensile | Uniaxial Load | 62 |

**Table 1.** Summary of modulus values of GC reported in the literature.

Further, as discussed in Albiez and Schwaiger [39], the modulus of GC seems to be affected by geometry and precursor material. This is consistent with the insights developed by MD modeling reported here as well as a previous work [9] where the formation of complex and amorphous GC structures is significantly dependent on such effects like substrate (more dominant for thinner films), maximum pyrolysis temperature and ramping rate [5], as well as annealing. Taken together, this suggests that some of the large variations in the reported modulus could be accounted for by these factors. It is also important to mention that for GC structures supported on a polyimide substrate, the net modulus of the composite shows the significant effect of the modulus of polyimide layer which is in the order of 2.5 GPa [35]. Similar results were reported by our group where the modulus of a composite structure consisting of two layers of polyimide (HD 4100), metal traces, and a final layer of stiffer durimide 7520 was found to be ~2.35 GPa [7], which is again comparable to the modulus of durimide 7520 which is 2.5 GPa.

## 4. CONCLUSIONS
The work presented here provides modeling approach towards investigating the mechanical properties of GC using molecular dynamics simulations which model the atomistic level formation and breaking of bonds using bond-order based reactive force field formulations. The MD modeling offered some unique insights that will contribute to our understanding of the mechanical property of GC. These are:

1. GC has uniquely different behavior under tensile and compressive loadings due to its complex and amorphous nature. The flat graphene-like component with its *6-member* rings and $sp^2$ hybridization is less stiff than the 3D cagey structure and forms the locus of failure where bonds break progressively. On the other hand, the 3D cagey structure which is dominated by 5-*and 7-member* rings offers high resistance to both tensile and compressive loads.
2. The failure mode in GC under uniaxial tensile loading is highly influenced by $sp^2$ bonds that eventually break and reduce to *c-c* chains (*sp* bond), as failed bonds transfer load to the adjacent atomic bonds.
3. The stress-strain relationships in GC demonstrate several peaks post the first failure, consistent with work reported for graphene and graphene oxide [27-28].
4. Increase in temperature and loading strain rates do not seem to affect the modulus and elastic response of GC but have noticeable effects in post-failure regime.

On the other hand, the mechanical uniaxial tensile load tests that were carried out to provide validation as well as insights to the bulk behavior of GC demonstrated the following:
1. Failure in bare GC structures is dominated by brittle-type behavior arising from the amorphous nature of GC much like what its name suggests.
2. The bulk material behavior of GC is similar to its molecular behavior in predicting a brittle tensile failure; however, the modulus of the bulk material is significantly higher than that of the molecular model due to the complex structure of the final bulk material that is made of millions of GC molecules with defects, impurities and pores.
3. The modulus of GC reported here is consistent with those reported in the literature; however, the modulus is affected by the maximum pyrolysis temperature (carbonization level), process parameters used like gas flow rate (internal pores) and thickness that could in turn impact the structure of the final GC bulk structure.
4. The presence of a polymer substrate supporting the GC structure helps in creating a composite structure that exhibits a ductile failure mode. This outcome is consistent with results reported before where the failure of the final GC structure was dominated by a tearing type of failure plane at the ultimate stress. This demonstrates that GC supported on flexible substrate offers an excellent pathway for a mechanically robust platform for using GC in such applications like *in vivo* implantations.

The MD molecular-level modeling and the mechanical uniaxial tensile test presented here, taken together, offer new insights that improve our understanding of the tensile and compressive behavior of GC. However, future work in modeling its behavior under shear, for example, could add more to this in addition to the consideration of several units of GC molecules within the same simulation cell.

**Acknowledgment**
This material is partially based on research work supported by the Center for Neurotechnology (CNT), a National Science Foundation Engineering Research Center (EEC-1028725).

**Competing Financial Interests:**
The authors declare no competing financial interests.

**Author contributions: M.K** built the simulation models, carried out the simulations, analyzed the results and wrote the materials and methods and results sections of the paper. **L.C.C** carried out the fabrication and characterized the test structures, and also wrote the materials and methods and results sections of the paper; **S.N**. carried out the initial simulations and interpretation of results; **J.B.** helped with the plots; **M.E.V, D.T.R, C.F**, and **S.S** helped in simulations and fabrications; **B.K.C**. helped with plots; **S.K.** formulated the concept, supervised the project, structured the outline of the paper, edited the manuscript, and wrote the introduction, discussion and conclusion section of the paper.